\documentclass[aip, cha, reprint]{revtex4-2}
\usepackage{xspace,units}
\usepackage{graphicx, multirow}
\usepackage{amssymb, amsmath, amsfonts}
\usepackage{mathtools}
\usepackage{tabularx}
\usepackage{dcolumn}% Align table columns on decimal point
\usepackage{bm}% bold math
\usepackage{enumitem}
\usepackage[utf8]{inputenc}
\usepackage[T1]{fontenc}

\begin{document}

\title{Ordinal methods for a characterization of evolving functional brain networks}
\author{Klaus Lehnertz}
\email{klaus.lehnertz@ukbonn.de}
\affiliation{Department of Epileptology, University of Bonn Medical Centre, Venusberg Campus 1, 53127 Bonn, Germany}
\affiliation{Helmholtz Institute for Radiation and Nuclear Physics, University of Bonn, Nussallee 14--16, 53115 Bonn, Germany}
\affiliation{Interdisciplinary Center for Complex Systems, University of Bonn, Br{\"u}hler Stra\ss{}e 7, 53175 Bonn, Germany}

\begin{abstract}
Ordinal time series analysis is based on the idea to map time series to ordinal patterns, i.e., order relations between the values of a time series and not the values themselves, as introduced in 2002 by C. Bandt and B. Pompe.
Despite a resulting loss of information, this approach captures meaningful information about the temporal structure of the underlying system dynamics as well as about properties of interactions between coupled systems.
This --~together with its conceptual simplicity and robustness against measurement noise~-- makes ordinal time series analysis well suited to improve characterization of the still poorly understood spatial–temporal dynamics of the human brain.
This minireview briefly summarizes the state-of-the-art of uni- and bivariate ordinal time-series-analysis techniques together with applications in the neurosciences.
It will highlight current limitations to stimulate further developments which would be necessary to advance characterization of evolving functional brain networks.
\end{abstract}
\maketitle

\begin{quotation}
Deriving evolving functional brain networks from observed, long-lasting, multivariate time series to improve characterization of various physiological and pathophysiological brain dynamics requires suitable and robust time-series-analysis techniques, that are capable of deciphering the multifaceted nature of the brain's complex endogenous and exogenous interactions.
I will recapitulate concepts of ordinal time series analysis, showcase its applications in the neurosciences, and will discuss limitations and necessary developments to improve characterization of the complex networked dynamics system human brain.
\end{quotation}

%%%%%%%%%%%%%%%%%%%%%%%%%%%%%%%%%%%%%%%%%%%%%%%%%%%%%%%%%%%%%%%%%%%%%%%%%%%%%%%%%%%%%%%%%%
\section{Introduction}
\label{sec:intro}
\begin{figure}[h]
\begin{centering}
 	\includegraphics[width=1.\columnwidth]{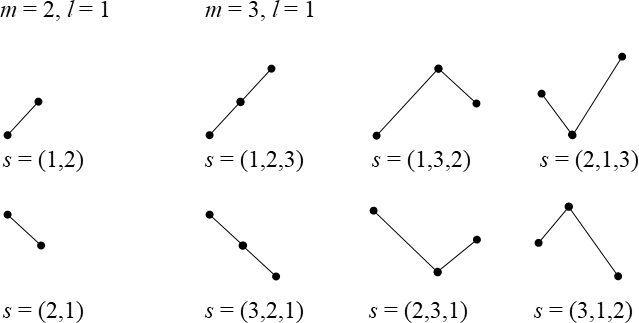}
	\end{centering}
	\caption{Possible outcomes for ordinal patterns (permutation symbols) $s$ using embedding dimensions $m=2$ and $m=3$ and fixed embedding delay $l=1$.
	}
	\label{fig:fig1}
\end{figure}

%%%%%%%%%%%%%%%%%%%%%%%%%	%%%%%%%%%%%%%%%%%%%%%%%%%%%%%%%%%%%%%%%%%%%%%%%%%%%%%%%%%%%%%%%%%
%
Ordinal time series analysis is a special type of symbolic analysis~\cite{hao1989,daw2003} which makes use of \textit{symbols} that are ordinal patterns (also referred to as order patterns or permutation patterns) of length of at least 2.
C. Bandt and B. Pompe introduced this concept in 2002 in their seminal paper~\cite{bandt2002} together with permutation entropy as a natural complexity measure of time series. 
Let $x_i=x(i),i=1,\ldots,N$, denote a sequence of observations (or time series) from some system $X$. 
For a given, but otherwise arbitrary $i$, $m$ amplitude values $X_i=\{x(i),x(i+l),\ldots,x(i+(m-1)l)\}$ are arranged in an ascending order $\{x(i+(k_{i1}-1)l)\leq x(i+(k_{i2}-1)l)\leq\ldots\leq x(i+(k_{im}-1)l)\}$, where $l$ and $m$ denote the appropriately chosen~\cite{staniek2007,Berger2019,Myers2020} time delay and embedding dimension (cf. Takens' embedding theorem~\cite{takens1981,sauer1991}). 
In case of equal amplitude values, one can e.g. carry out the rearrangement according to the associated index $k$, i.e., for $x(i+(k_{i1}-1)l)=x(i+(k_{i2}-1)l)$ one can write $x(i+(k_{i1}-1)l)\leq x(i+(k_{i2}-1)l)$ if $k_{i1}<k_{i2}$ thereby ensuring that every $X_i$ is uniquely mapped onto one of the $m!$ possible permutations. 
A permutation symbol --~or ordinal pattern~-- is then defined as $s_i \equiv(k_{i1}, k_{i2}, \ldots, k_{im})$ and captures qualitative information about the temporal structure of the underlying time series (see Fig~\ref{fig:fig1}).

Ordinal time series analysis is conceptually simple, computationally fast and comparably robust against measurement noise. 
Compared to other symbolization techniques~\cite{daw2003}, the derivation of ordinal patterns does not require a priori knowledge about the data range, which rendered ordinal time series analysis beneficial for investigations of empirical data from various scientific domains~\cite{Amigo2010,Zanin2012b,Amigo2013,Unakafova2013,Amigo2015,Keller2017,Keller2019,Piek2019,Gutjahr2020,Zanin2021,Leyva2022}.
A large proportion of studies was concerned with problems such as distinguishing chaos from noise, improving the detection of determinism~\cite{Amigo2010a,Small2018,Hirata2019} or of dynamical changes~\cite{cao2004}, system identification~\cite{Parlitz2013},
or quantifying time reversibility~\cite{Zanin2018a}, thereby employing ordinal-pattern-derived quantifier for entropy~\cite{Piek2019,Gutjahr2020}, complexity~\cite{Unakafova2013}, or combinations thereof~\cite{Rosso2007}.

In this minireview, we will concentrate on ordinal time-series-analysis techniques that aim at characterizing properties of interactions --~strength, direction, and coupling function~--, since these currently form the basis of complex-network-based studies in diverse scientific fields including geophysics, meteorology, and the neurosciences~\cite{boccaletti2006,arenas2008,barthelemy2011,holme2012,bassettsporns2017,Gosak2018,Halu2019,Wang2022}. 
With this ansatz, one assumes that a spatially extended complex system can be represented by a complex network which, however, requires identification of vertices and edges.
In many cases, such an identification is straightforward, but it remains a challenging issue when investigating the system's dynamics~\cite{ioannides2007,butts2009,bialonski2010,hlinka2012,papo2016,hlinka2017,Korhonen2021,Rings2022}.
Network vertices are usually assumed to represent distinct subsystems and edges represent interactions between them, and these vertices and edges constitute a functional (or interaction) network.
In an evolving functional network, properties of edges (and/or vertices) are time-dependent~\cite{lehnertz2014}.
In case that a direct access to interactions and their time-dependencies is not possible (e.g. via probing), one usually resorts to linear and non-linear time-series-analysis techniques to quantify interaction properties from pairs of time series of appropriate system observables. 
These techniques originate from diverse fields such as statistics, synchronization theory, non-linear dynamics, information theory, statistical physics, and from the theory of stochastic processes~\cite{pikovsky2001,boccaletti2002,kantz2003,pereda2005,hlavackova2007,marwan2007,lehnertz2011,stankovski2017,Runge2018,gorjao2019,tabar2019book,Papana2021,Papana2021a}, given that interactions can manifest themselves in various aspects of the dynamics.
While the majority of studies on (evolving) functional networks is based on binary (an edge exists or not) or weighted networks (the weight of an edge is given by the strength of interaction), further improvements can be expected by considering weighted \textit{and} directed networks, thereby including knowledge about coupling functions that contain detailed information about the functional mechanisms underlying an interaction and that prescribe the physical rule specifying how an interactions occurs~\cite{Stankovski2019}.

%%%%%%%%%%%%%%%%%%%%%%%%%%%%%%%%%%%%%%%%%%%%%%%%%%%%%%%%%%%%%%%%%%%%%%%%%%%%%%%%%%%%%%%%%%
\section{Ordinal Methods for a characterization of interactions}
\label{sec:inter}
%%%%%%%%%%%%%%%%%%%%%%%%%%%%%%%%%%%%%%%%%%%%%%%%%%%%%%%%%%%%%%%%%%%%%%%%%%%%%%%%%%%%%%%%%%
%
Current bivariate ordinal time-series-analysis techniques allow characterization of strength and direction of interactions; whether some of these techniques also allow for a characterization of coupling functions needs further investigations. 

\begin{figure}
\begin{centering}
 	\includegraphics[width=1.\columnwidth]{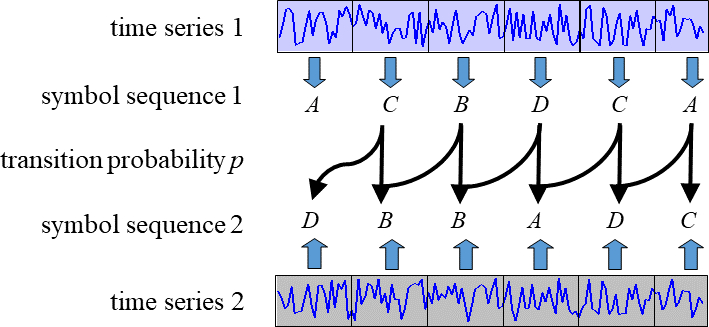}
	\end{centering}
	\caption{
	Schematic of time-resolved ordinal-pattern-based estimation of direction of interactions. 
	Sequences of permutation symbols (here: $A, B, C$, and $D$) derived from successive segments of time series (moving-window technique).
	Direction of interactions can be quantified by estimating the differential flow of information between symbol sequences (from sequence 1 to sequence 2 and vice versa) via transition probabilities $p$ between current and past states (black arrows). 
	}
	\label{fig:fig2}
\end{figure}

Estimators for the strength of interactions center around the phenomenon of synchronization and its various forms of appearance --~from complete via phase and lag synchronization to generalized synchronization~\cite{pikovsky2001,Boccaletti2018}).
In case of generalized synchronization, the relationship between two empirical time series can be characterized by an order parameter that is based on the consistent changing tendency of their permutation entropies~\cite{liu2004} (see Fig.~\ref{fig:fig2}). 
The value of the order parameter can be used to assess the strength of an interaction.
In case of phase synchronization, the latter can be assessed with a metric (so-called ordinal synchronization) that is based on the dot product between two ordinal vectors~\cite{Echegoyen2019}.
Analysis techniques based on transcripts~\cite{monetti2009} as well as those based on conceptual extensions (so-called coupling complexity)~\cite{amigo2012,Monetti2013b} allow one to detect and characterize various forms of synchronization and the strength of an interaction can be assessed with different derived estimators.

The majority of estimators for the direction of interactions are based on the information-theoretic functional conditional mutual information~\cite{Cover2006,Bossomaier2016} and need to be defined asymmetrically to allow detection of a directed flow of information  (see Fig.~\ref{fig:fig2}).
Among these estimators are directionality indices based on the so-called PI approach~\cite{bahraminasab2008}, on symbolic transfer entropy~\cite{staniek2008,staniek2009,kugiumtzis2012,dickten2014,martini2011}, 
on momentary information transfer~\cite{pompe2011}, on transcripts~\cite{monetti2013,Amigo2016}, on closeness mapping~\cite{Amigo2018}, on the ratio of the number of missing joint ordinal patterns~\cite{Yin2019}, or on joined symbolic recurrences~\cite{Porfiri2019}. 
Some of these estimators take into account coupling delays~\cite{pompe2011,dickten2014}, which is of importance as it allows for improved physical interpretations.
There are also extensions that enable the time-resolved investigation of directional relationships between coupled dynamical systems from short and transient noisy time series~\cite{martini2011}.
With respect to complex-network-based studies, an estimator for the direction of interactions should allow for distinguishing direct from indirect directional couplings, as this is a key to avoid severe misinterpretations of possible causal relationships.
So far, only one of the aforementioned approaches (symbolic transfer entropy) has been extended using partialization analysis to allow for such a distinction~\cite{kugiumtzis2013b,Papana2016}.
It remains to be investigated, however, whether these extensions suffer from limitations similar to the ones identified for other partialized estimators for the direction of interactions when investigating interactions in larger networks (number of nodes $\gg 10$)~\cite{rings-lehnertz2016}.

I briefly mention complementary approach from the field of time series networks~\cite{Zou2019}, i.e., a transformation of a time series into the complex network domain, namely transition networks derived from ordinal patterns as well as cross and joint 
ordinal transition networks. 
These networks can then be investigated further to characterize strength and direction of interactions~\cite{Zhang2017ordpat,Guo2018,Zou2019,Ruan2019,Subramaniyam2021}.
There are, however, some issues that would need to be fully resolved to allow judging general applicability of these approaches.

%%%%%%%%%%%%%%%%%%%%%%%%%%%%%%%%%%%%%%%%%%%%%%%%%%%%%%%%%%%%%%%%%%%%%%%%%%%%%%%%%%%%%%%%%%
\section{Towards characterizing evolving functional brain networks with ordinal methods}
\label{sec:brain}
%%%%%%%%%%%%%%%%%%%%%%%%%%%%%%%%%%%%%%%%%%%%%%%%%%%%%%%%%%%%%%%%%%%%%%%%%%%%%%%%%%%%%%%%%%
%
Ordinal methods appear to be ideally suited to improve characterization of the complex time-varying dynamics of the human brain~\cite{lehnertz2017} given its contrasting and, at times, complicated forms of appearance: oscillations at a variety of frequencies~\cite{Niedermeyer2004} coexisting with scale-free dynamics ($1/f^\alpha$-like power spectrum (with $\alpha \in \left[1,3\right]$) at many spatial-temporal scales~\cite{Gisiger2001,bedard2006,Marom2010,Werner2010,Buzsaki2012,He2014}).
Indeed, a number of studies provided evidence for univariate ordinal time-series-analysis techniques to detect spatial-temporal changes of EEG data related to epileptic activity in a quantitative and efficient way that may provide helpful information for diagnostic and therapeutic purposes~\cite{Keller2003,cao2004,Keller2005,Keller2007,Ouyang2010,schindler2011,schindler2012,Rummel2013,Keller2014,Redelico2017,Zeng2018,Granado2022}.
Similar observations were made for other neurological and neurodevelopmental disorders such as traumatic brain injury~\cite{Kalpakis2015}, mild cognitive impairment and Alzheimer’s disease~\cite{Morabito2012}, Parkinson's disease~\cite{Yi2017ordpat} and attention-deficit/hyperactivity disorder~\cite{Amigo2010b}, and all these disorders impose a high individual, medical, psychosocial and socioeconomic burden for those affected.
The techniques were also shown to allow differentiating sleep states~\cite{Nicolaou2011,Keller2014,Bandt2017,Gonzalez2019,Hou2021} and various awake states (including different mental activities)~\cite{Dimitriadis2012b,Dimitriadis2016,QuinteroQuiroz2018}, brain dynamics during resting states related to different age groups~\cite{Kottlarz2021}, as well as brain states related to anesthesia~\cite{Jordan2008,Olofsen2008,Li2010b,Sarasso2021}.

Regardless of the success of univariate ordinal time-series-analysis techniques, one should keep in mind that the majority of physiological and pathophysiological changes in brain dynamics investigated so far are accompanied by distinct modifications of the respective brain dynamics' frequency content which --~in many cases~-- is visible to the naked eye.
Future studies should thus demonstrate advantages of univariate ordinal time-series-analysis techniques over \textit{classical} techniques~\cite{Berger2017,YamashitaRiosdeSousa2022}, also to increase their acceptance for clinical applications. 

In contrast to the aforementioned rather large number of possible applications of univariate ordinal time-series-analysis techniques, there are so far only a few studies that employed bivariate ordinal techniques to characterize interactions between various pairs of brain regions.
Although mainly restricted to multichannel recordings from subjects with epilepsy, techniques allowed a comparably thorough characterization of strength~\cite{liu2004} and direction~\cite{staniek2008} of short- to long-ranged interactions between structurally identical and nonidentical coupled but not yet fully synchronized brain regions --~covering lobes and hemispheres~-- from time-resolved analyses of multiday recordings that captured a large variety of physiological and pathophysiological brain states~\cite{osterhage2007,staniek2007,staniek2008,staniek2009,dickten2014,lehnertz-dickten2015}.
These investigations also provided evidence for bivariate ordinal time-series-analysis techniques to allow an improved characterization of brain interaction dynamics that can be regarded predictive of an impeding epileptic seizure~\cite{lehnertz-dickten2015}, a prerequisite for the development of refined seizure prevention or control techniques~\cite{kuhlmann2018}.
Nevertheless, these studies also identified potential limitations if confounding variables such as delayed interactions, asymmetric signal-to-noise ratios, number of interacting systems or
connection densities are not taken into account.  
In general, these studies suggested to estimate both strength and direction of interactions in order to effectively distinguish various coupling regimes (uncoupled, weak to strong couplings)
and to avoid misinterpretations when investigating directional interactions between complex dynamical systems~\cite{lehnertz-dickten2015}.

Bivariate ordinal time-series-analysis techniques were also shown to reliably detect and characterize changes of transient (in the order of a few 100 milliseconds) directional interactions between brain regions associated with cognitive control~\cite{martini2011} as well as to identify topographical reorganizations of interactions between brain regions related to coma~\cite{Zubler2016}, stroke~\cite{Zubler2018}, or 	anesthesia-induced unconsciousness~\cite{Lee2018}. 

By now, investigations of evolving functional brain networks employing bivariate ordinal time-series-analysis techniques are rare and exclusively related to epilepsy~\cite{zubler2015,dickten2016}.
A highly time-resolved investigation of importance of vertices in evolving functional brain networks during a large number of seizures identified ``hub''-like brain regions that appear to be of relevance for the complex spatial-temporal spreading dynamics~\cite{zubler2015}.
Interestingly, these brain regions only rarely coincided with the clinically defined seizure onset zone, which calls for revisiting the role of the latter in seizure generation~\cite{geier2015,dickten2016,rings2019precursors,Fruengel2020}. 
Eventually, a comparative study employed bivariate ordinal~\citep{liu2004,staniek2008} and phase-synchronization-based~\citep{mormann2000,rosenblum2001a} time-series-analysis techniques to characterize weighted and directed interactions in functional brain networks --~that evolve over several days~-- from a large group of subjects with epilepsy~\cite{dickten2016}.
On a population-sample level and despite the heterogeneity of investigated cases, both approaches appeared to provide comparable information about the network characteristics.
On the level of individual subjects, however, the approaches provided largely independent, non-redundant information but with a varying contrast. 
This can probably be related to the various concepts (different synchronization phenomena, information flow) from which time-series-analysis techniques were derived.

%%%%%%%%%%%%%%%%%%%%%%%%%%%%%%%%%%%%%%%%%%%%%%%%%%%%%%%%%%%%%%%%%%%%%%%%%%%%%%%%%%%%%%%%%%
\section{The next steps}
\label{sec:next}
%%%%%%%%%%%%%%%%%%%%%%%%%%%%%%%%%%%%%%%%%%%%%%%%%%%%%%%%%%%%%%%%%%%%%%%%%%%%%%%%%%%%%%%%%%
%
Summarizing the state-of-the-art of bivariate ordinal time-series-analysis techniques in use to characterize complex interactions, it can be noted that there are more estimators for the direction of an interaction than for the strength, notwithstanding the coupling function. 
While estimators for the strength of interactions were mainly designed to capture the various forms of synchronization, the majority of estimators for the direction are based on the information-theoretic functional conditional mutual information, for which many approximations are available.
Given this imbalance, many studies in the neurosciences (as well as in other disciplines) employed ordinal time-series-analysis techniques to either estimate the direction or the strength of interactions, or estimated both but by employing ordinal techniques that were derived from different concepts.
In the latter case, it is important to note that the techniques' efficiency may be influenced differently by a number of confounding factors: 
volume conduction effects~\cite{albo2004,vinck2011}, 
propagation delays and delayed couplings~\cite{silchenko2010,dickten2014,Govindan2006}, 
asymmetric signal-to-noise ratios~\cite{nolte2004,nolte2008,Vinck2015} or eigenfrequency ratios~\cite{nolte2008,Wu2008} (in case of oscillating (sub)systems), 
peculiarities of the recording~\cite{guevara2005,zaveri2006,Hu2011,porz2014,SanzGarcia2018,Snyder2018,RiosHerrera2019}, 
pre-processing steps such as filtering~\cite{Florin2011,Gerster2022}, 
the techniques' capability to distinguish between (apparent) interdependencies due to common sources and (true) interdependencies due to interacting (sub)systems~\cite{nolte2008,porz2014}, the techniques' capability to distinguish between direct and indirect interactions~\cite{nawrath2010,mader2015,rings-lehnertz2016,Lu2022}, or the techniques' different sensitivities for the various types of synchronization phenomena, to name just a few.
Many confounding factors can be identified by investigating e.g. coupled paradigmatic model systems with well-known properties. 
The impact of such factors can be minimized by improving the techniques' robustness and by using surrogate concepts~\citep{schreiber2000a,ansmann2012,stahn2017}, but these are rarely used in the context of ordinal time-series-analysis techniques. 
More importantly, we still lack reliable surrogate concepts to probe for the direction of interactions. 

An exception to the aforementioned conceptual mixture are transcripts-based estimators for strength and direction of interactions~\citep{monetti2009,monetti2013}, which share a common conceptual base.  
So far, their general applicability and particularly for the neurosciences has been demonstrated only by way of example, and their susceptibility to the aforementioned (and other) confounding factors remains to be elucidated.

Future extensions and/or improvements of estimators for strength and direction of pairwise and higher-order\citep{Battiston2020} interactions, of coupling functions as well as of concepts underlying ordinal methods can be expected to further increase capability of ordinal time-series-analysis techniques to investigate time-evolving networks --~in the neurosciences as well as in other scientific domains concerned with networked dynamical systems.

%%%%%%%%%%%%%%%%%%%%%%%%%%%%%%%%%%%%%%%%%%%%%%%%%%%%%%%%%%%%%%%%%%%%%%%%%%%%%%%%%%%%%%%%%%
\section{Conclusions}
\label{sec:concl}
%%%%%%%%%%%%%%%%%%%%%%%%%%%%%%%%%%%%%%%%%%%%%%%%%%%%%%%%%%%%%%%%%%%%%%%%%%%%%%%%%%%%%%%%%%

In this paper, I have briefly reviewed the state-of-the-art of univariate and bivariate ordinal time-series-analysis techniques thereby focusing on applications in the neurosciences.
I also discussed current limitations to stimulate further developments which would be necessary to advance characterization of evolving functional brain networks during both physiological and pathophysiological conditions.
Ordinal time series analysis carries the potential to improve characterization of the still poorly understood spatial–temporal dynamics of the human brain.

\section*{Acknowledgements}
In memory of our esteemed colleague Karsten Keller, who unexpectedly passed away in April 2022.

\section*{Data availability}
Data sharing is not applicable to this article as no new data were created or analyzed in this study.

\section*{References}

\end{document}